# Islanding, growth mode and ordering in Si heteroepitaxy on Ge(001) substrates structured by thermal annealing

L. Persichetti,[1,*] M. Fanfoni,[2] B. Bonanni,[2] M. De Seta,[1] L. Di Gaspare,[1] C. Goletti,[2] L. Ottaviano,[3] and A. Sgarlata[2]

[1]*Dipartimento di Scienze, Università Roma Tre, Viale G. Marconi, 446- 00146 Roma, Italy*

[2]*Dipartimento di Fisica, Università di Roma "Tor Vergata", Via Della Ricerca Scientifica, 1- 00133 Roma, Italy*

[3]*Dipartimento di Scienze Fisiche e Chimiche, Università degli Studi dell'Aquila, L'Aquila, Italy*

Si/Ge heteroepitaxial dots under tensile strain are grown on nanostructured Ge substrates produced by high-temperature flash heating exploiting the spontaneous faceting of the Ge(001) surface close to the onset of surface melting. A very diverse growth mode is obtained depending on the specific atomic structure and step density of nearby surface domains with different vicinal crystallographic orientations. On highly-miscut areas of the Ge(001) substrate, the critical thickness for islanding is lowered to about 5 ML, in contrast to the 11 ML reported for the flat Ge(001) surface, while on unreconstructed (1$x$1) domains the growth is Volmer-Weber driven. An explanation is proposed considering the diverse relative contributions of step and surface energies on misoriented substrates. In addition, we show that the bottom-up pattern of the substrate naturally formed by thermal annealing determines a spatial correlation for the dot sites.

**I. INTRODUCTION**

Strain is ubiquitous in heteroepitaxial growth. Strain control and engineering have been the focus of intense research in the last decades, with the optimization of self-organization schemes based on the strain-driven Stranski-Krastanow (SK) growth mode being a central topic [1-3]. By changing the aspect ratio and shape of SK quantum dots, one can tune, respectively, the magnitude and the symmetry of the strain field [4, 5], whereas the sign of the strain is fixed by the relative values of the lattice parameters in the dot and substrate. As a matter of fact, the paradigm of SK growth is a compressively strained island grown on a wetting layer (WL), also being under lattice compression over the substrate. This is the case for the two model systems of SK heteroepitaxy, i.e. Ge/Si(001) [3, 6-10] or InAs/GaAs(001) [11, 12]. Indeed, growing the reverse tensile-strained SK interface where the dots are under tensile strain, such as Si/Ge(001), is much more problematic and the number of studies on the SK Si/Ge(001) growth is rather limited [13-18], as compared to the huge amount of works on the Ge/Si(001) SK growth. The main hurdle is that a tensile-strained 2D Si/Ge(001) film is significantly more stable against islanding than a wetting layer under compression [19], despite the main driving force for islanding, i.e. the elastic energy, scaling as $\varepsilon^2$, is independent of the strain sign. Accordingly, at a growth temperature $T= 500$ °C, the onset of three-dimensional growth has been reported to exceed 11 monolayers (MLs) in Si/Ge(001) [15, 16], while it is only 3-4 MLs in Ge/Si(001) both for the flat [20-22] and vicinal orientations [23-25]. Such asymmetry has been largely debated in literature and diverse explanations have been proposed in terms of strain-dependent diffusivity [26], step [19] and surface energies [27]. However, identifying the main driving force is still an open issue. Even more puzzling is that, on the Ge(111) surface, a SiGe wetting layer is instead not observed since the Si/Ge(111) growth proceeds in the Volmer-Weber (VW) mode [28, 29].

In this paper, we investigate the growth of Si dots on naturally structured Ge(001) substrates obtained by high-temperature flash heating. Exploiting the spontaneous faceting of the Ge(001) surface close to the onset of surface melting [30], we compare the stability against islanding of tensile-strained Si/Ge wetting layers under the same growth conditions but on domains with different vicinal crystallographic orientations close to the Ge(001) face. These domains include staircases of highly-miscut $(2x1)$-reconstructed (001) terraces separated by a dense array of steps, flat (115)-faceted surfaces with the $c(3x1)$ reconstruction and disordered and unreconstructed areas showing the (1x1) bulk termination. The substrate morphology is therefore ideal to directly assess the relative contributions of step and surface energies to the stability of planar tensile-strained films. We find that, on this structured Ge surface, the 2D-to-3D transition during Si overgrowth occurs at markedly lower thickness than for the flat Ge(001) surface. On the highly-miscut Ge(001) domains, the critical WL thickness is 5 ML while on the unreconstructed areas the growth is VW driven, showing that the specific surface structure has a critical effect on the strain-induced islanding under lattice tension. In addition, we show that the nucleation sites of the Si dots are spatially correlated to the faceting-pattern periodicity, thus suggesting a viable bottom-up self-assembly route.

## II. MATERIALS AND METHODS

Experiments were carried out by using commercial epi-ready, prime-grade polished Czochralski (CZ) Ge(001) wafers (N-type, Sb-doped and resistivity of about 5-7 Ω cm). The accuracy on the polar orientation specified by the supplier was ±0.5°, whereas on the azimuthal orientation was better than ±2°. Sample preparation was performed in ultra-high-vacuum (UHV) conditions ($p< 5x10^{-11}$ mbar) by several cleaning cycles consisting of Ar$^+$ sputtering (830 V, 20 min) and annealing to 1103 K by direct current heating. As checked by scanning tunnelling microscopy (STM), the surface shows, at this stage, the typical morphology of the flat Ge(001) face, with large (001) terraces having alternating $(2x1)$ and $(1x2)$ reconstructions [31]. The faceting pattern is obtained by direct-current flash heating to $T \gtrsim 1173$ K for few seconds followed by a rapid thermal quench to room temperature obtained by abruptly interrupting the heating current passing through the sample. Being close to the Ge melting point, temperature calibration is critical and was performed by optical pyrometry on a sacrificial melted sample. We also carefully checked that no appreciable temperature inhomogeneities were present on the sample, the colour of which was uniform during annealing. *In situ* STM measurement were performed at room-temperature using W tips and in constant-current mode. Si was deposited at $T=$ 500 °C by physical vapor deposition (PVD) from an ultra-pure rod using an electron-beam evaporator directly pointing at the STM stage. In this evaporation geometry, a Si flux ranging between 0.01 and 0.6 ML/min could be achieved. In order to accurately calibrate the Si deposition rate, we perform, in the same experimental conditions, a test experiment in which the rate was obtained from real-time STM measurements acquired during step-flow growth in Si/Si(001) homoepitaxy [32].

## III. RESULTS AND DISCUSSION

*High-temperature structure of the Ge(001) surface*

Figure 1 shows the morphology of the Ge(001) surface after the high-temperature annealing step to $T \gtrsim$ 1173 K. Consistently with previous reports [30], the surface structure is completely altered with respect to the flat (2*x*1)-(1*x*2) terraces obtained for low-temperature annealing [33, 34]. From large-scale STM images [Fig. 1(a, b)], we observe the formation of a superstructure pattern with periodicity $w=$ (36 ± 5) nm coexisting with disordered areas of the sample. A blow-up [Fig. 1(c)] reveals that the superstructure consists of a faceted domain where two crystallographic faces are present: One face is a flat Ge(115) surface with the characteristic *c*(3*x*1) reconstruction [enlarged view shown in Fig. 1(d)] [35]; the other one is a high-miscut vicinal surface composed by a staircase of narrow (001) terraces with an average width of (1.5 ± 0.3) nm. As expected for large misorientation angles [31], high-resolution STM [Fig. 1(e)] shows that adjacent (001) terraces have the same dimerization direction, being parallel to the step ledge (*B*-type terraces), and that the step height is double (*D*) the mono-atomic Ge(001) plane. The small misalignment between the dimer rows and the average step direction which emerges from Fig. 1(e) is compatible with the uncertainty over the azimuthal orientation in the wafer cut which leads to the formation of forced kinks in the step profiles. Conversely, the disordered domains lack of any atomic reconstruction and appear rough at the nanoscale. Previous spectroscopic results indicate a substantial (2*x*1)-dimer breakup at high temperature and attribute surface disorder to the absence of in-plane bonding for the (1x1) bulk termination of Ge [36]. As a matter of fact, it is known that, for temperatures close to the melting point, Ge exhibits surface melting, i.e. the solid-vapour interface is wet by a mobile and wiggling *quasi-liquid* layer which forms at the surface [37, 38]. This means that the outermost layer is highly mobile and shows intermediate structural properties between a solid and a liquid. In our case, the observation of disordered patches, clearly reminiscent of a liquid, indicates that the final annealing step takes place above the onset of the Ge surface melting. We believe that, due to the rapid cooling, the quasi-liquid layer is quenched and the surface is frozen in an out-equilibrium morphology. Accordingly, the large misorientation found in the ordered domains agrees well with a frozen wiggling of the liquid layer. Moreover, in contrast to metals, the onset of surface melting in Ge occurs in close proximity to the melting point (*incomplete* melting) [37] and this explains why the structuring in Ge is only observed for *T*> 1173 K.

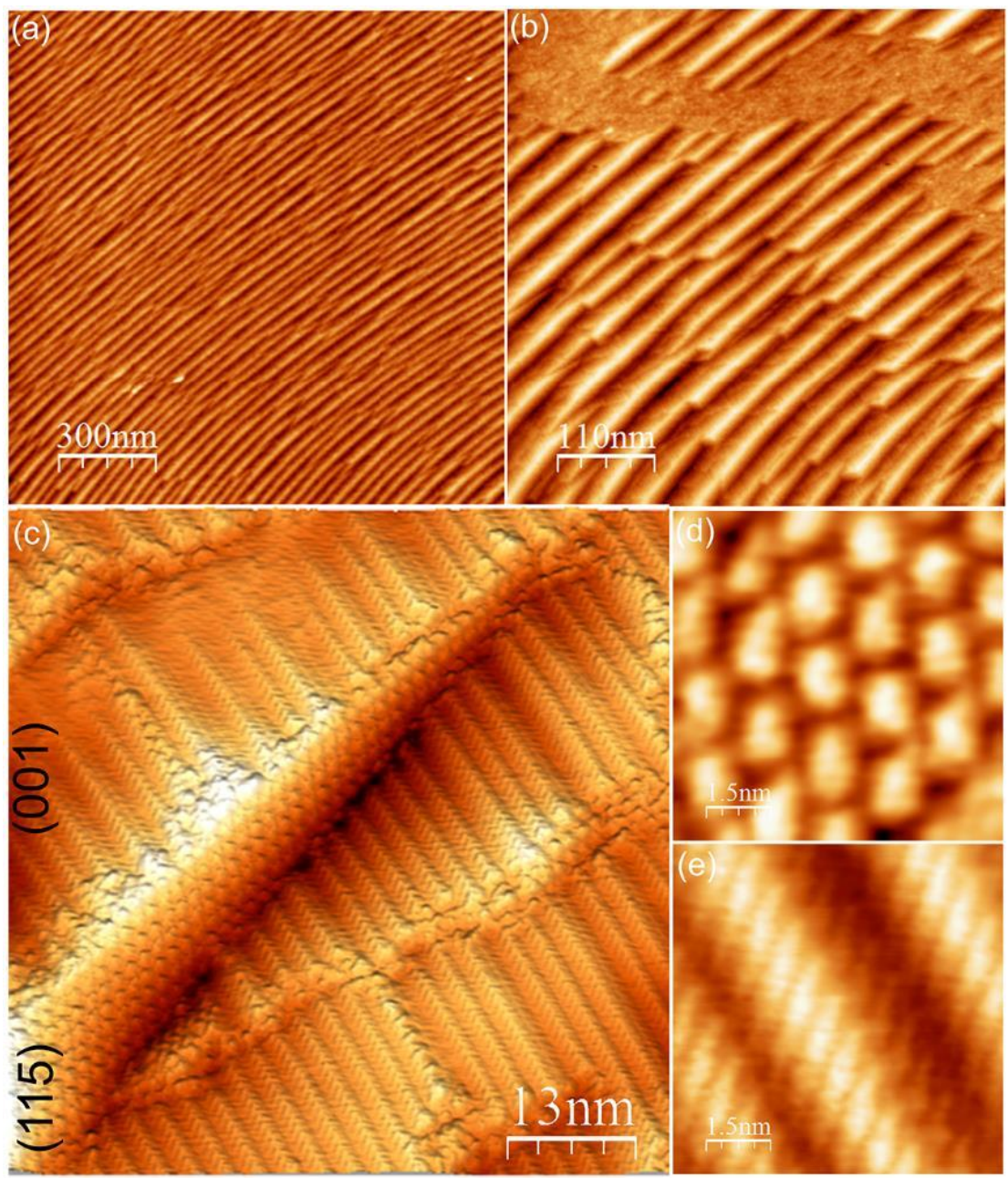

FIG. 1. STM images of the structured Ge(001) surface obtained by high-temperature flash annealing to 1173 K followed by rapid cooling. Ordered domains (a) coexist with disordered patches visible in (b). (c) Enlarged view of the ordered areas shows highly-miscut (001) terraces and (115) facets. (d, e) Blow-ups of (c) the *c*(3*x*1) reconstruction of the (115) facets and of (d) the (2*x*1) reconstruction of the (001) terraces. The diagonals of the panels are aligned along the <110> directions.

*Si/Ge heteroepitaxy*

We now follow by STM the nanoscale morphological evolution of this naturally structured Ge surface during Si overgrowth. For a Si coverage $\Theta$= 0.5 ML, the disordered domains already appear quite rugged [Fig. 2(a)], having a root-mean-square roughness $\sigma$= 3.4 Å. In fact, we observe the nucleation of a plenty of small 3D clusters with heights of 1-2 nm and lateral sizes below 10 nm. The onset of 3D growth at such low coverage indicates the occurrence of a VW process on the (1$x$1) domains. Conversely, at the same coverage, the high-miscut (2$x$1)-reconstructed Ge(001) regions [Fig. 2(b)] show lower roughness ($\sigma$= 0.82 Å), being well below 1 Å and increased only slightly from the value of 0.78 Å before Si overgrowth. By carefully inspecting Fig. 2(b), we note that surface dimers are asymmetric after Si deposition. This results in the appearance of a zigzag pattern of buckled dimer rows (indicated by the arrow) which have a *p*(2$x$2) in-phase arrangement one respect to the other [39]. In line with previous observations [40-42], the stabilization of asymmetric dimers at room temperature is a fingerprint of SiGe intermixing in the WL; mixed SiGe dimers store a significant amount of strain which is relieved in part by the buckling. The formation of an intermixed WL indicates that Si is preferentially incorporated into the 2D layer and does not form, at this stage, a 3D clusters, in contrast to the unreconstructed domains. We note that nucleation of clusters is also not observed on the Ge(115) facets [Fig. 2(a)].

At higher Si coverage $\Theta$= 4.1 ML [Fig. 2(c, d)], we still discern the tracks of the (115) facets; in between them, however, the dense array of (001) steps is not visible anymore but it is replaced by a continuous film. Being known that the energy of Si(001) steps is substantially increased under tensile strain [19], the high step density of the (001) domains evidently becomes energetically unfavourable at large WL thickness. The continuous film shows a non-negligible roughness ($\sigma$= 2.60 Å) and, locally, a few 3D clusters are already present. In Fig. 2(d), the highest cluster detected, having a height of 3 nm and a lateral size of 6 nm, is observed close to an edge of the faceting pattern which is clearly a preferential nucleation site. On the original surface topography before Si overgrowth, this location corresponds to a site on the (001) domain in close proximity to the edge of a nearby (115) facet. On the same sites, we observe the formation of large 3D islands when the Si coverage is slightly increased to $\Theta$= 5.5 ML [Figs. 2(e, f) and 3(a, b)] and, thus, we estimate 5 ML as the critical thickness for the 2D to 3D transition on the structured part of the surface. The islands have an average height of 10 nm and a lateral size of 30 nm. Information on the crystallographic orientations of island facets are obtained through the so-called facet plot (FP) analysis [43]. Each spot in the inset of Fig. 3(b) represents a different facet orientation with respect to the substrate plane being in the center of the diagram. From the FP, the main orientations of the island facets are the {113} (circles) and the {15 3 23} (squares), which are typically dominant for SiGe multifaceted islands [15, 44, 45].

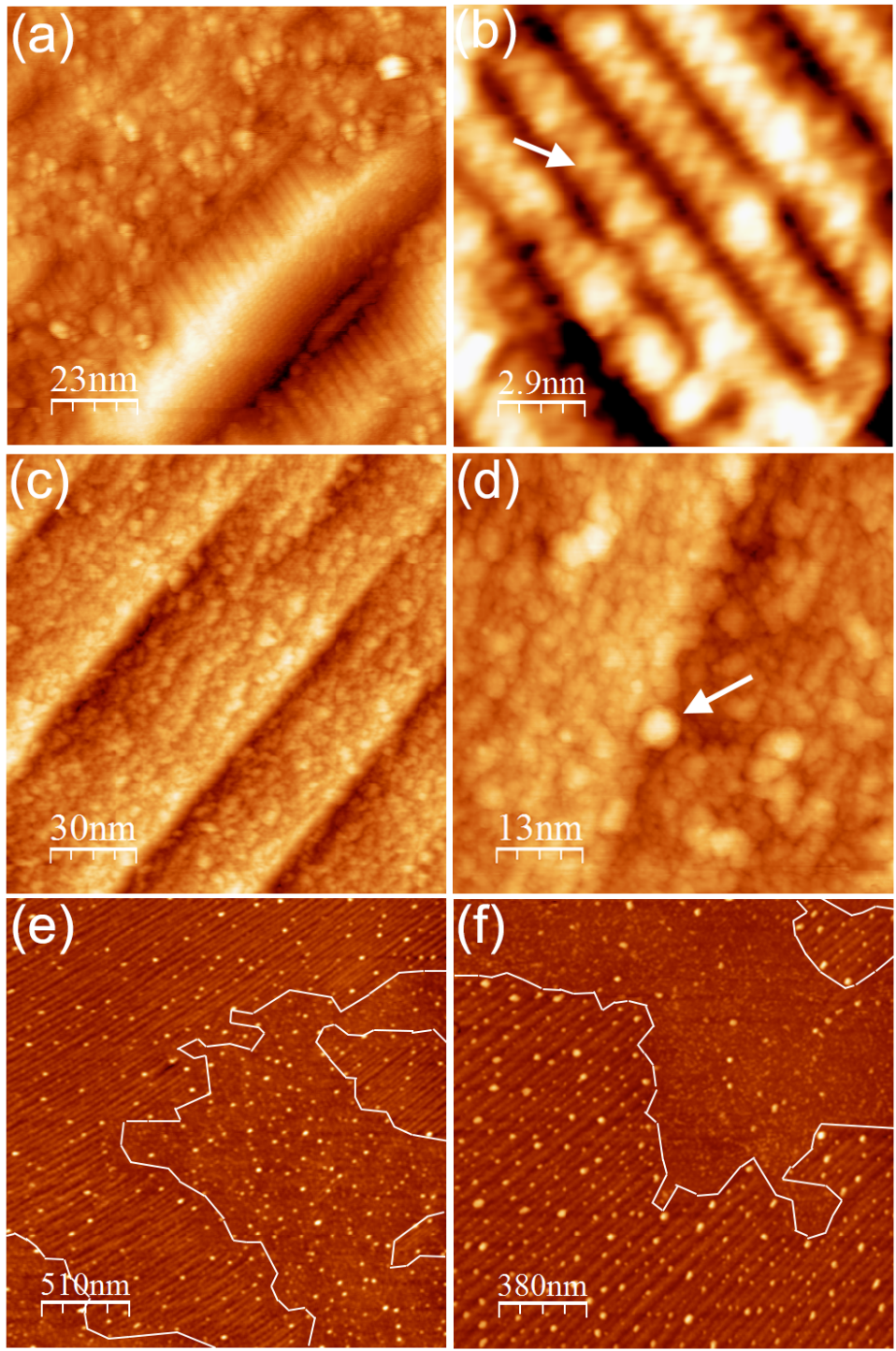

FIG. 2. STM images of the substrate at different Si coverage. (a, b) Θ= 0.5 ML. (c, d) Θ= 4.1 ML. (e, f) Θ= 5.5 ML. In panel (b), we observe the characteristic zigzag pattern of the *p*(2*x*2) reconstruction (indicated by the arrow). In panel (d), the arrow points to a 3D cluster nucleated along the faceting edge of the substrate. In the last two panels the disordered (1*x*1) patches are highlighted. The diagonals of the panels are aligned along the <110> directions.

*Growth-modes in Si/Ge heteroepitaxy*

We showed that the surface structure has a major impact on the Si/Ge growth mode, the critical thickness for 3D islanding being negligible on the unreconstructed (1$x$1) domains and about 5 ML on highly-stepped Ge(001), in contrast to the 11 ML-thick WL reported for the flat Ge(001) surface at the same growth temperature [15]. Besides, remember that the Si/Ge(111) growth has been reported to be of the VW type [28, 29].

Following Bauer's criterion [46], since the surface energy of Ge is smaller than that of Si (for relaxed unreconstructed (1$x$1) terminations and different surface reconstructions both on the (001) and on the (111) surfaces [47]), one would expect no WL formation for the Si/Ge interface. However, recent *ab-initio* calculations showed a subtle strain-dependence of surface energies which is completely different for the (001) and the (111) orientations [48, 49]. Namely, while tensile strain stabilizes the (2$x$1)-reconstructed Si(001) surface [48], the surface energy of the Si(111) surface is instead increased under tensile strain [49]. We believe that this different interplay is critical to explain the experimental observations. A coherent Si film grown on Ge shows a tensile strain $\varepsilon$= +4.2%; at this level of strain, the 2$x$1-Si(001) surface energy is markedly lowered and becomes comparable to that of the unstrained Ge(001) [48]. This means that the surface-energy gain favouring VW islanding no longer exists under tensile strain and it explains why planar Si growth is stabilized in Si/Ge(001) heteroepitaxy. Conversely, the surface energy of a Si(111) film under tensile strain on Ge(111) becomes even larger than for the unstrained surface [49]. Thus, for Si/Ge(111) the condition for VW is strengthened, well matching the experimental observations [28, 29]. The radically different behaviour of Si(111) is a further indication that the stabilization of the layer-by-layer growth is intimately related to the specific strain-dependent behaviour of the 2$x$1-reconstructed Si(001) surface. Thus, we do not expect any strain-induced reduction of surface energy for the unreconstructed (1$x$1) Si(001) domains where, in fact, we observe the inherent VW behaviour.

The only open question is now what drives the lower critical thickness for islanding observed on the highly-stepped (2$x$1) regions compared to the flat (001) surface. Since the planar growth is stabilized by the strain, any strain relaxation mechanism leads to an increase of the surface energy of the Si(001) overlayer and destabilizes the layer-by-layer growth. It is therefore interesting to consider what is the behaviour of surface stress for vicinal surfaces close to the (001) orientation of Si (or Ge). Two competing effects emerge as the step separation *L* decreases when the miscut angle gets larger. It is known that, since the presence of steps leads to surface stress relaxation, the surface stress is maximum for the low-index flat surface, and decreases with the miscut angle, scaling as ∼*L* [50, 51]. On this overall decreasing trend, another effect arises at low miscut angles, which is the transition between single- and double-stepped surfaces, occurring for Ge(001) around 4° miscut [31]. Since the single-layer steps are always accompanied by alternating (1$x$2) and (2$x$1) terraces with rotated surface stress, there is an additional contribution to surface relaxation which reduces the surface stress for single-stepped surfaces at low miscuts and scales with the terrace width as $\sim ln(L/\pi a)$, where $a \approx 0.4$ nm is the surface lattice constant [31, 52]. In our case, the highly-stepped structured regions have a very small step separation corresponding to a miscut angle larger than 10°. At this angle, the surface stress reduction due to the reduced coordination at the step sites is dominant and the surface stress is surely reduced with respect to the flat surface [50, 51], thus increasing surface energy of the overgrown film. We remark, however, that at lower miscut angles where the surface relaxation due to surface stress anisotropy (i.e. the logarithmic term above) is effective, the balance of the two aforementioned contributions is expected to be delicate and a more complex growth behaviour may arise. For the highly-miscut (2x1) domains we observe, in addition, the energetics of steps becomes non-negligible, due the higher relative contribution of step-energy to the total surface energy [53]. It is known that the step energy of Si(001) is strongly increased by tensile strain [19] and this implies that forming a coherent Si film reproducing the stepped morphology of the misoriented Ge(001) underneath has a high energy cost. Consistently, we found that, for sufficiently thick Si films, the stepped structure is lost. We propose that the combination of surface stress reduction and step edge energy, which are related to the dense step array observed on our naturally structured Ge(001) substrate, lowers the stability of a 2D film and drives the formation of islands at lower critical thickness than on the flat Ge(001) surface.

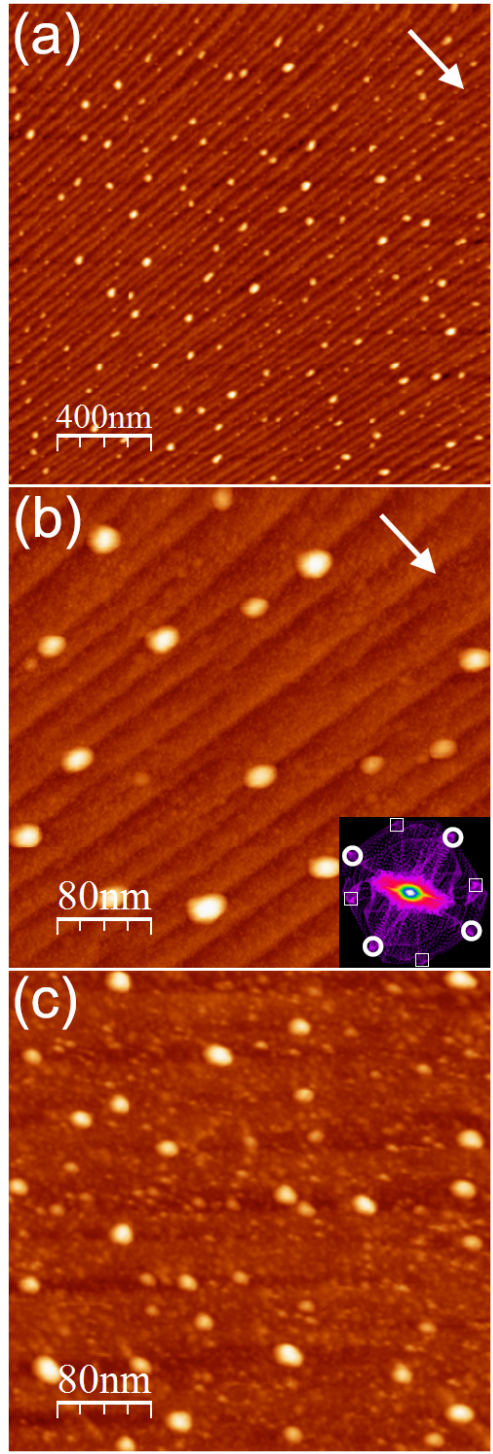

FIG. 3. STM images of the Si islands grown at $\Theta= 5.5$ ML on (a, b) the structured domains and (c) on the unstructured domains. In the inset of panel (b), the island's FP is reported. {113} facets are marked by circles, while the {15 3 23} by squares. The arrows indicate the $[1\bar{1}0]$ direction.

*Spatial ordering of Si/Ge islands induced by surface structuring*

In Figs. 3(a) and 3(b), there is a clear correlation between the island's ordering and the structuring pattern of the substrate. As a matter of fact, while in the unstructured areas the islands appear homogenously distributed [Fig. 3(c)], a distinct spatial ordering is present in the spontaneously structured domains [Fig. 3(a, b)]. Confirming what observed at lower coverage, large 3D islands occupy selectively specific locations on the structured areas: They sit on the WL grown on the (001) domains and, specifically, are aligned along the edge boundaries with the (115) facets [Fig. 3(b)]. On the surface pattern, these edges are oriented along the [110] direction and are evenly spaced with a period $w$; we, therefore, expect to find some quantitative correlation between $w$ and the island's separation along the [1$\bar{1}$0]. In Fig. 4(a), we compare the distribution of the nearest-neighbour distances projected along the [1$\bar{1}$0] for an island ensemble grown on a structured portion of the substrate (upper histogram) with that obtained on the unstructured areas (lower histogram). As expected, the latter appears structureless, consistently with the absence of texture in the substrate. Conversely, on the spontaneously structured areas of the substrate, the distribution clearly shows distinct peaks which we labelled as $L_n$, with $n= 0 \div 4$. The peaks are evenly spaced with a period $L_1=$ 37 nm well matching the pattern periodicity $w$. It is easy to see that, if the islands are aligned along the edges of the pattern, as drawn in Fig. 4(b), the projection $Lx$ of the island separation along the [1$\bar{1}$0] only assumes values which are multiples of the pattern periodicity $w$. Thus, we interpret the zero-order peak $L_0$ as due to the islands having the nearest-neighbour on the same faceting edge, and the following orders as due to islands with the nearest-neighbour at increasingly distant locations on the pattern. Confirming the visual impression from the STM images, this peak structure is a quantitative manifestation of the ordering effect driven by the substrate pattern. The latter also affects the size distribution of the Si dots which is clearly narrower on the structured domains [Fig. 3(a, b)] with respect to the unstructured areas [Fig. 3(c)]. In line with previous observations [54, 55], we interpret this effect as the result of the preferentially unidimensional mass transport occurring along the faceting edges in the structured areas. In contrast, on the unstructured areas, the diffusion is entirely two-dimensional and this, together with the early nucleation of small clusters at low coverage, results in broader island's size distribution.

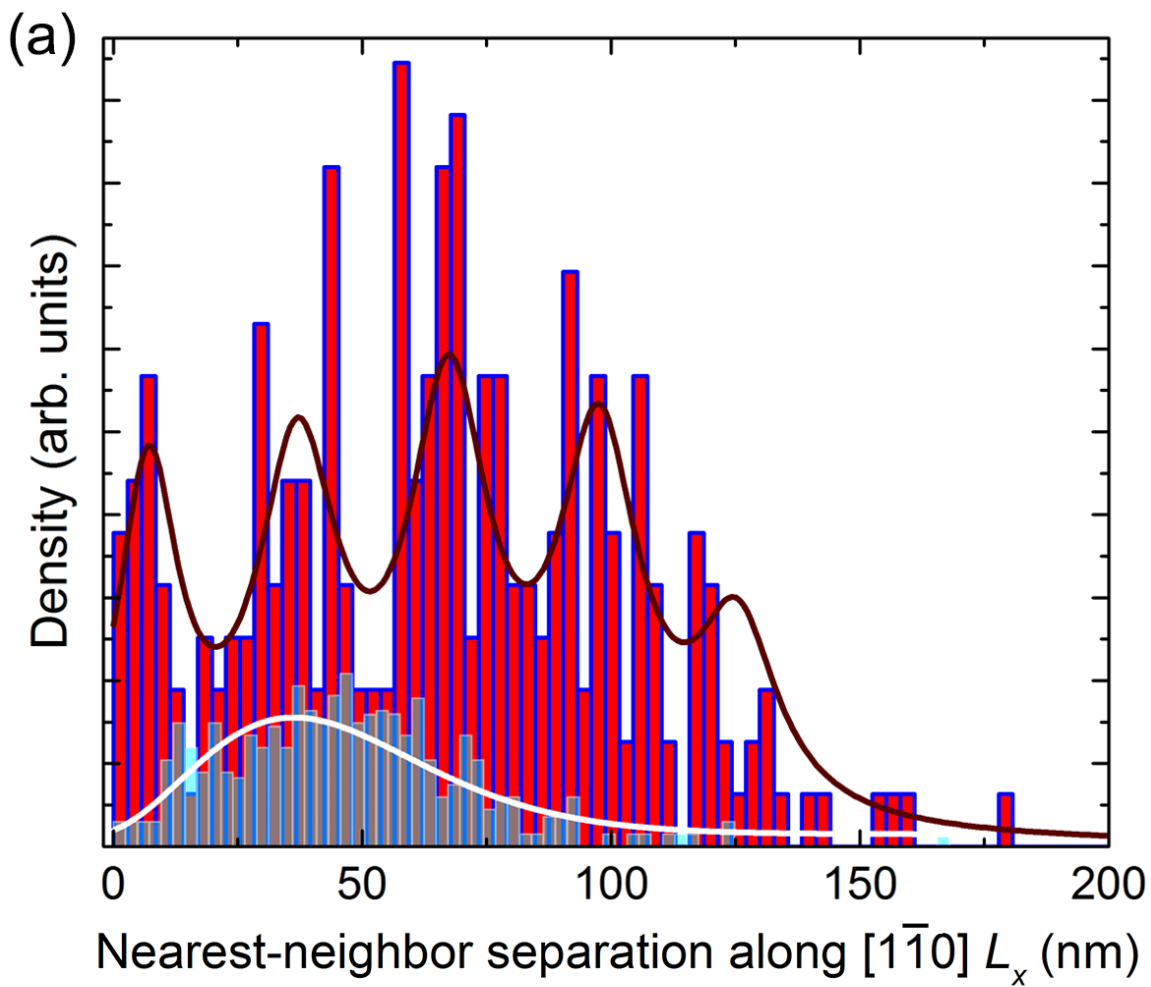

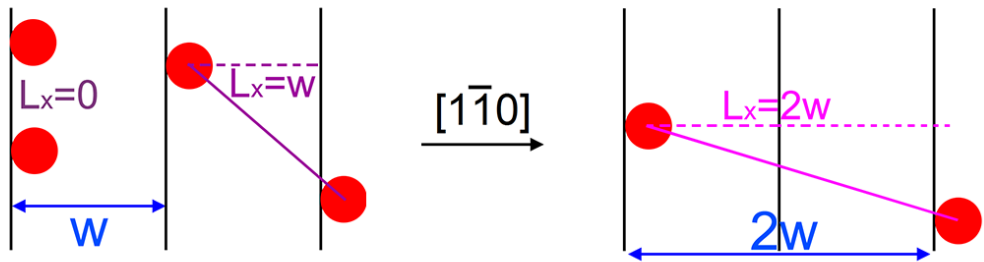

FIG. 4. (a) Distribution of the projections of nearest-neighbour distances along the $[1\bar{1}0]$ direction (i.e. the direction perpendicular to the edges of the faceting pattern) for Si islands at $\Theta$= 5.5 ML. The upper histogram plot shows the distribution on the spontaneously structured domains, whereas the lower one on the unstructured areas. For improving readability, the vertical scale of the latter has been divided by a factor 5. (b) Schematics of possible island pair configurations on the surface pattern with periodicity $w$. The corresponding $L_x$ is reported in each case.

**IV. CONCLUSIONS**

We investigated the Si/Ge growth on naturally structured Ge substrates obtained by high-temperature flash heating exploiting the spontaneous faceting of the Ge(001) surface close to the onset of surface melting. The formation of surface domains with different vicinal crystallographic orientations made it possible to compare, in the same growth conditions, the stability against islanding of tensile-strained Si/Ge wetting layers on templates showing diverse relative contributions of step and surface energies. We found that on a Ge(001) substrate with high step density the critical thickness for islanding is lowered to about 5 ML, in contrast to the 11 ML reported on the flat Ge(001) surface; while on the unreconstructed (1$x$1) domains of the pattern the growth is VW driven, showing that the specific surface structure critically affects strain-induced islanding under lattice tension. In addition, we show that the substrate structure defines a spatial correlation for nucleation sites which resembles the faceting-pattern periodicity, thus suggesting a doable bottom-up strategy for self-assembly of Si dots.

**REFERENCES**

[1] J. Stangl, V. Holý, G. Bauer, Structural properties of self-organized semiconductor nanostructures, Reviews of Modern Physics, 76 (2004) 725-783.

[2] I. Berbezier, A. Ronda, SiGe nanostructures, Surf. Sci. Rep., 64 (2009) 47-98.

[3] J.N. Aqua, I. Berbezier, L. Favre, T. Frisch, A. Ronda, Growth and self-organization of SiGe nanostructures, Phys. Rep., 522 (2013) 59-189.

[4] L. Persichetti, A. Sgarlata, M. Fanfoni, A. Balzarotti, Breaking Elastic Field Symmetry with Substrate Vicinality, Phys. Rev. Lett., 106 (2011) 055503.

[5] L.A.B. Marçal, M.I. Richard, L. Persichetti, V. Favre-Nicolin, H. Renevier, M. Fanfoni, A. Sgarlata, T.Ü. Schülli, A. Malachias, Modified strain and elastic energy behavior of Ge islands formed on high-miscut Si(0 0 1) substrates, Appl. Surf. Sci., 466 (2019) 801-807.

[6] G. Capellini, M. De Seta, F. Evangelisti, Influence of the growth parameters on self-assembled Ge islands on Si(100), Mater. Sci. Eng: B, 89 (2002) 184-187.

[7] M. De Seta, G. Capellini, L. Di Gaspare, F. Evangelisti, F. D'Acapito, Freezing shape and composition of Ge∕Si(001) self-assembled islands during silicon capping, J. Appl. Phys., 100 (2006) 093516.

[8] L. Persichetti, A. Sgarlata, M. Fanfoni, A. Balzarotti, Heteroepitaxy of Ge on singular and vicinal Si surfaces: elastic field symmetry and nanostructure growth, J. Phys.: Condens. Matter, 27 (2015) 253001.

[9] F. Montalenti, F. Rovaris, R. Bergamaschini, L. Miglio, M. Salvalaglio, G. Isella, F. Isa, H. von Känel, Dislocation-Free SiGe/Si Heterostructures, Crystals, 8 (2018).

[10] L. Fazi, C. Hogan, L. Persichetti, C. Goletti, M. Palummo, A. Sgarlata, A. Balzarotti, Intermixing and buried interfacial structure in strained Ge/Si(105) facets, Phys. Rev. B, 88 (2013) 195312.

[11] E. Placidi, F. Arciprete, R. Magri, M. Rosini, A. Vinattieri, L. Cavigli, M. Gurioli, E. Giovine, L. Persichetti, M. Fanfoni, F. Patella, A. Balzarotti, InAs Epitaxy on GaAs(001): A Model Case of Strain-Driven Self-assembling of Quantum Dots Self-Assembly of Nanostructures, in: S. Bellucci (Ed.) Self-Assembly of Nanostructures, Springer New York, 2012, pp. 73-125, doi: 10.1007/978-1-4614-0742-3_2.

[12] J. Wu, P. Jin, Self-assembly of InAs quantum dots on GaAs(001) by molecular beam epitaxy, Frontiers of Physics, 10 (2015) 7-58.

[13] W. Wulfhekel, H.J.W. Zandvliet, B.-J. Hattink, G. Rosenfeld, G. Comsa, B. Poelsema, Smooth growth fronts in Si/Ge heteroepitaxy by kinetic growth manipulation, Phys. Rev. B, 58 (1998) 15359-15362.

[14] H.J.W. Zandvliet, E. Zoethout, W. Wulfhekel, B. Poelsema, Origin of roughening in epitaxial growth of silicon on Si(001) and Ge(001) surfaces, Surf. Sci., 482-485 (2001) 391-395.

[15] D. Pachinger, H. Groiss, H. Lichtenberger, J. Stangl, G. Hesser, F. Schäffler, Stranski-Krastanow growth of tensile strained Si islands on Ge (001), Appl. Phys. Lett., 91 (2007) 233106.

[16] D. Pachinger, H. Groiss, M. Teuchtmann, G. Hesser, F. Schäffler, Surfactant-mediated Si quantum dot formation on Ge(001), Appl. Phys. Lett., 98 (2011) 223104.

[17] W.H. Tu, C.H. Lee, H.T. Chang, B.H. Lin, C.H. Hsu, S.W. Lee, C.W. Liu, A transition of three to two dimensional Si growth on Ge (100) substrate, J. Appl. Phys., 112 (2012) 126101.

[18] L. Persichetti, A. Sgarlata, S. Mori, M. Notarianni, V. Cherubini, M. Fanfoni, N. Motta, A. Balzarotti, Beneficial defects: exploiting the intrinsic polishing-induced wafer roughness for the catalyst-free growth of Ge in-plane nanowires, Nanoscale Research Letters, 9 (2014) 358.

[19] Y.H. Xie, G.H. Gilmer, C. Roland, P.J. Silverman, S.K. Buratto, J.Y. Cheng, E.A. Fitzgerald, A.R. Kortan, S. Schuppler, M.A. Marcus, P.H. Citrin, Semiconductor Surface Roughness: Dependence on Sign and Magnitude of Bulk Strain, Phys. Rev. Lett., 73 (1994) 3006-3009.

[20] P.D. Szkutnik, A. Sgarlata, S. Nufris, N. Motta, A. Balzarotti, Real-time scanning tunneling microscopy observation of the evolution of Ge quantum dots on nanopatterned Si(001) surfaces, Phys. Rev. B, 69 (2004) 201309.

[21] M. Brehm, F. Montalenti, M. Grydlik, G. Vastola, H. Lichtenberger, N. Hrauda, M. Beck, T. Fromherz, F. Schäffler, L. Miglio, G. Bauer, Key role of the wetting layer in revealing the hidden path of Ge/Si(001) Stranski-Krastanow growth onset, Phys. Rev. B, 80 (2009) 205321.

[22] R. Bergamaschini, M. Brehm, M. Grydlik, T. Fromherz, G. Bauer, F. Montalenti, Temperature-dependent evolution of the wetting layer thickness during Ge deposition on Si(001), Nanotechnol., 22 (2011) 285704.

[23] L. Persichetti, A. Sgarlata, M. Fanfoni, A. Balzarotti, Shaping Ge Islands on Si(001) Surfaces with Misorientation Angle, Phys. Rev. Lett., 104 (2010) 036104.

[24] L. Persichetti, A. Sgarlata, M. Fanfoni, A. Balzarotti, Ripple-to-dome transition: The growth evolution of Ge on vicinal Si(1 1 10) surface, Phys. Rev. B, 82 (2010) 121309.

[25] L. Persichetti, A. Sgarlata, M. Fanfoni, A. Balzarotti, Pair interaction between Ge islands on vicinal Si(001) surfaces, Phys. Rev. B, 81 (2010) 113409.

[26] B.J. Spencer, P.W. Voorhees, J. Tersoff, Morphological instability theory for strained alloy film growth: The effect of compositional stresses and species-dependent surface mobilities on ripple formation during epitaxial film deposition, Phys. Rev. B, 64 (2001) 235318.

[27] J. Tersoff, Step Energies and Roughening of Strained Layers, Phys. Rev. Lett., 74 (1995) 4962-4962.

[28] P.M.J. Marée, K. Nakagawa, F.M. Mulders, J.F. van der Veen, K.L. Kavanagh, Thin epitaxial Ge−Si(111) films: Study and control of morphology, Surf. Sci., 191 (1987) 305-328.

[29] A. Raviswaran, C.-P. Liu, J. Kim, D.G. Cahill, J.M. Gibson, Evolution of coherent islands during strained-layer Volmer-Weber growth of Si on Ge(111), Phys. Rev. B, 63 (2001) 125314.

[30] L. Persichetti, A. Sgarlata, M. Fanfoni, A. Balzarotti, Irreversible order-disorder transformation of Ge(0 0 1) probed by scanning tunnelling microscopy, J. Phys.: Condens. Matter, 27 (2015) 435001.

[31] H.J.W. Zandvliet, The Ge(001) surface, Phys. Rep., 388 (2003) 1-40.

[32] B. Voigtländer, Fundamental processes in Si/Si and Ge/Si epitaxy studied by scanning tunneling microscopy during growth, Surf. Sci. Rep., 43 (2001) 127-254.

[33] J.S. Hovis, R.J. Hamers, C.M. Greenlief, Preparation of clean and atomically flat germanium(001) surfaces, Surf. Sci., 440 (1999) L815-L819.

[34] W.M. Klesse, G. Scappucci, G. Capellini, M.Y. Simmons, Preparation of the Ge(001) surface towards fabrication of atomic-scale germanium devices, Nanotechnol., 22 (2011) 145604.

[35] Z. Gai, R.G. Zhao, X. Li, W.S. Yang, Faceting and nanoscale faceting of Ge{hhl} surfaces around (113), Phys. Rev. B, 58 (1998) 4572-4578.

[36] E. van Vroonhoven, H.J.W. Zandvliet, B. Poelsema, (2x1)-(1x1) Phase Transition on Ge(001): Dimer Breakup and Surface Roughening, Phys. Rev. Lett., 91 (2003) 116102.

[37] U. Tartaglino, T. Zykova-Timan, F. Ercolessi, E. Tosatti, Melting and nonmelting of solid surfaces and nanosystems, Phys. Rep., 411 (2005) 291-321.

[38] A.D. Laine, M. DeSeta, C. Cepek, S. Vandré, A. Goldoni, N. Franco, J. Avila, M.C. Asensio, M. Sancrotti, Surface phase transitions of Ge(100) from temperature-dependent valence-band photoemission, Phys. Rev. B, 57 (1998) 14654-14657.

[39] R. Wolkow, Direct observation of an increase in buckled dimers on Si(001) at low temperature, Phys. Rev. Lett., 68 (1992) 2636-2639.

[40] X. Chen, D.K. Saldin, E.L. Bullock, L. Patthey, L.S.O. Johansson, J. Tani, T. Abukawa, S. Kono, Atomic geometry of mixed Ge-Si dimers in the initial-stage growth of Ge on Si(001)2x1, Phys. Rev. B, 55 (1997) R7319-R7322.

[41] X.R. Qin, B.S. Swartzentruber, M.G. Lagally, Scanning Tunneling Microscopy Identification of Atomic-Scale Intermixing on Si(100) at Submonolayer Ge Coverages, Phys. Rev. Lett., 84 (2000) 4645-4648.

[42] P.D. Szkutnik, A. Sgarlata, A. Balzarotti, N. Motta, A. Ronda, I. Berbezier, Early stage of Ge growth on Si(001) vicinal surfaces with an 8° miscut along [110], Phys. Rev. B, 75 (2007) 033305.

[43] A. Rastelli, H. von Känel, Surface evolution of faceted islands, Surf. Sci., 515 (2002) L493.

[44] L. Persichetti, A. Capasso, S. Ruffell, A. Sgarlata, M. Fanfoni, N. Motta, A. Balzarotti, Ordering of Ge islands on Si(001) substrates patterned by nanoindentation, Thin Solid Films, 519 (2011) 4207-4211.

[45] R. Bergamaschini, J. Tersoff, Y. Tu, J.J. Zhang, G. Bauer, F. Montalenti, Anomalous Smoothing Preceding Island Formation During Growth on Patterned Substrates, Phys. Rev. Lett., 109 (2012) 156101.

[46] E. Bauer, Phanomenologische Theorie der Kristallabscheidung an Oberflachen. I, Zeitschrift fur Kristallographie, 110 (1958) 372-394.

[47] A.A. Stekolnikov, J. Furthmüller, F. Bechstedt, Absolute surface energies of group-IV semiconductors: Dependence on orientation and reconstruction, Phys. Rev. B, 65 (2002) 115318.

[48] C.V. Ciobanu, R.M. Briggs, Stability of strained H:Si(105) and H:Ge(105) surfaces, Appl. Phys. Lett., 88 (2006) 133125.

[49] R. Zhachuk, J. Coutinho, A. Dolbak, V. Cherepanov, B. Voigtländer, Si(111) strained layers on Ge(111): Evidence for c(2x4) domains, Phys. Rev. B, 96 (2017) 085401.

[50] J.J. Métois, A. Saúl, P. Müller, Measuring the surface stress polar dependence, Nat. Mater., 4 (2005) 238.

[51] P. Müller, A. Saùl, F. Leroy, Simple views on surface stress and surface energy concepts, Advances in Natural Sciences: Nanoscience and Nanotechnology, 5 (2013) 013002.

[52] O.L. Alerhand, D. Vanderbilt, R. Meade, J.D. Joannopoulos, Spontaneous Formation of Stress Domains on Crystal Surfaces, Phys. Rev. Lett., 61 (1988) 1973-1976.

[53] L. Persichetti, A. Sgarlata, G. Mattoni, M. Fanfoni, A. Balzarotti, Orientational phase diagram of the epitaxially strained Si(001): Evidence of a singular (105) face, Phys. Rev. B, 85 (2012) 195314.

[54] A. Sgarlata, L. Persichetti, A. Capasso, M. Fanfoni, N. Motta, A. Balzarotti, Driving Ge Island Ordering on Nanostructured Si surfaces, Nanoscience and Nanotechnology Letters, (2011) 841-849.

[55] E. Placidi, F. Arciprete, V. Latini, S. Latini, R. Magri, M. Scuderi, G. Nicotra, F. Patella, Manipulating surface diffusion and elastic interactions to obtain quantum dot multilayer arrangements over different length scales, Appl. Phys. Lett., 105 (2014) 111905.